\documentclass[11pt]{elsart}
\usepackage{amsfonts}
\usepackage{slashbox}

\usepackage{amssymb,amsmath}
\usepackage{mathrsfs} 
\usepackage{graphicx}

\begin{document}

\title{Signaling games with pattern recognition}
\author{Haoyang Wu}

\author{Department of Physics}

\author{Xi'an Jiaotong University, China.}

\author{hywch@mail.xjtu.edu.cn}

\textbf{Abstract}

The classical model of signaling games assumes that the receiver
exactly know the type space (private information) of the sender and
be able to discriminate each type of the sender distinctly. However,
the justification of this assumption is questionable. It is more
reasonable to let the receiver recognize the pattern of the sender.
In this paper, we investigate what happens if the assumption is
relaxed. A framework of signaling games with pattern recognition and
an example are given.



\section{Introduction}
Since the pioneering work of Spence\cite{Spence1973}, signaling
games have been deeply investigated by many researchers around the
world [2-10]. Generally, there are two players in the model, a
Sender (S) and a Receiver (R). The Sender has a certain type $t$ (an
element of a finite set $T$) which represents his private
information. The Sender observes his own type while the Receiver
does not know the type of the Sender. There is a strictly positive
probability distribution $p(t)$ on $T$: $p(t)$ represents the prior
probability that the Sender is of type $t$ and is the common
knowledge.

Based on his knowledge of his own type, the Sender chooses to send a
message from a set of possible messages $M=\{m_{1},\cdots,m_{J}\}$.
The Receiver observes the message but not the type of the Sender.
Then the Receiver chooses an action from a set of feasible actions
$A=\{a_{1},\cdots,a_{L}\}$. The two players receive payoffs
dependent on the Sender's type, the message chosen by the Sender and
the action chosen by the Receiver.

Although the aforementioned model has been widely used by the
economic community, people seldom notice clearly that they have made
an assumption in the model, i.e., the Receiver is able to
discriminate each type of the Sender distinctly. Actually, only by
this assumption can the Receiver be able to have a probability
distribution on the type set $T$. However, the justification of this
assumption is questionable. Because the type space $T$ is the
private information of the Sender, the Sender has no incentive to
tell his secret to the Receiver. The Receiver cannot take it for
granted that he can discriminate each type of the Sender exactly.

The aim of this paper is to investigate what happens if the
aforementioned assumption is canceled. We claim that what the
Receiver can do is to recognize the ``pattern'' of the Sender (The
phrase ``pattern recognition'' comes from computer science). A
pattern consists of one type or some types, and one type just
belongs to a pattern. The rest of the paper is organized as follows:
Section 2 discusses the model of signaling games with pattern
recognition. In Section 3, an example is given in detail.

\section{The model}
Without loss of generality, consider two players in a signaling game
with pattern recognition, $i=1,2$. Player 1 is the Sender who has
private information (i.e., the type). Player 2 is the Receiver.
Different from classical signaling games (where the Receiver is
assumed to know the type space of the sender and discriminate each
type exactly), here we relax this assumption and let the Receiver
recognize the pattern of the sender (A message can be sent from
several patterns. A pattern can send several messages). The pattern
space and the belief distribution hold by the Receiver are common
knowledge. The dynamic timing sequence of a signaling game with
pattern recognition is as follows:

Step 1: Nature ($N$) selects the type $t$ of player 1, $t\in
T=\{t_{1},\cdots,t_{K}\}$, $K\geq2$. Player 1 knows $t$, but player
2 doesn't know it. Furthermore, player 2 cannot discriminate each
type in $T$ exactly. The pattern space hold by player 2 is
$T'=\{t'_{1},\cdots,t'_{\bar{K}}\}$, $1\leq \bar{K}\leq K$. $T'$ is
a partition of $T$. The belief distribution hold by player 2 is
defined on the pattern space $T'$: $p(t')>0$, $t'\in T'$,
$\sum_{t'\in T'}p(t')=1$. A pattern $t'$ consists of some types, and
one type belongs to a pattern.

Step 2: After observing the type $t$, player 1 selects a message $m$
from his message space $M=\{m_{1},\cdots,m_{J}\}$.

Step 3: After receiving the message $m$ from player 1, player 2
makes an induction $p(t'|m)$ using Bayesian rules, and selects an
action $a$ from his feasible action space
$A=\{a_{1},\cdots,a_{L}\}$.

\textbf{Definition 1:} The payoffs of player 1 and player 2 are both
related to the type $t$ of player 1, the message $m$ and the action
$a$ of player 2, i.e., $u_{1}(t,m,a)$ and $u_{2}(t,m,a)$.

Note: 1) The strategy $s_{1}$ of player 1 is a message profile,
which is related to the type $t$ of player 1 and the pattern space
$T'$ hold by player 2, i.e., $s_{1}=(m_{11},\cdots,m_{1K})$, where
$m_{1k}(k,T')\in M=\{m_{1},\cdots,m_{J}\}$, $k=1,\cdots,K$.

2) The strategy $s_{2}$ of player 2 is an action profile, which is
related to the message $m$ and the pattern space $T'$ hold by player
2, i.e., $s_{2}=(a_{21},\cdots,a_{2J})$, where $a_{2j}(m,T')\in
A=\{a_{1},\cdots,a_{L}\}$, $j=1,\cdots,J$.

\begin{figure}
\centering
\includegraphics[height=2in,clip,keepaspectratio]{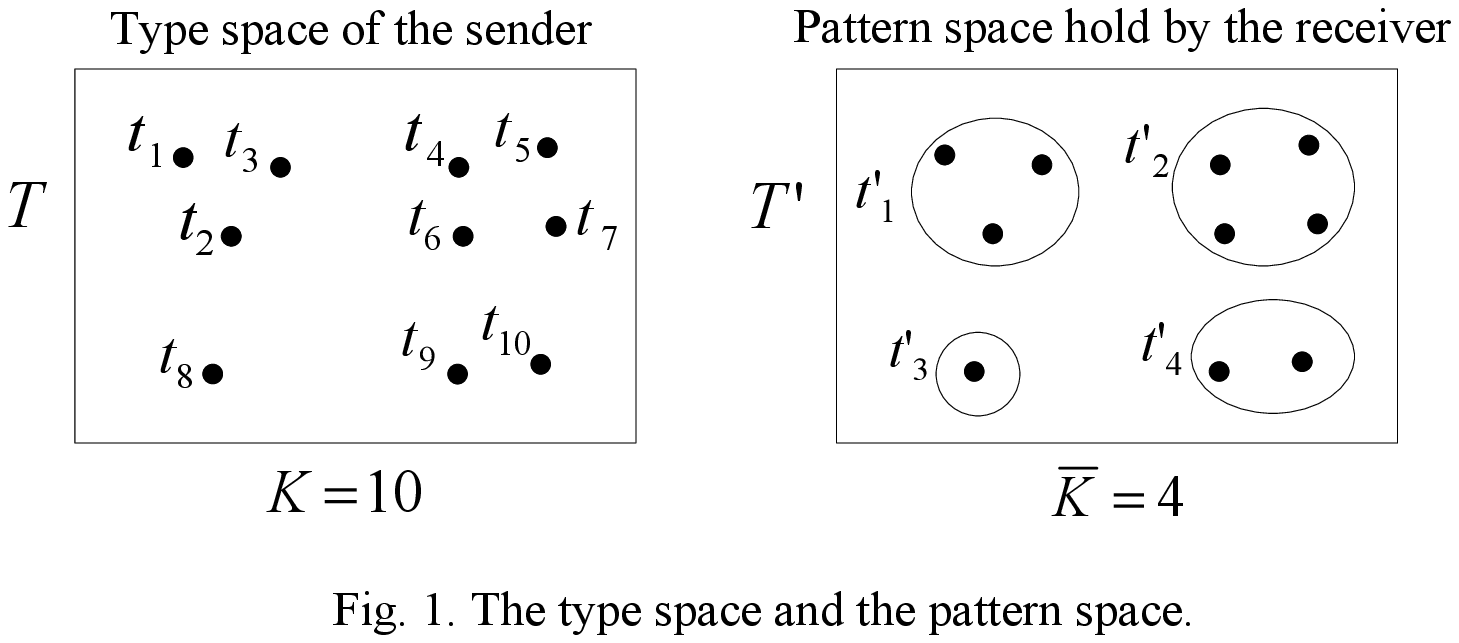}
\end{figure}

As shown in Fig. 1, $K=10$, $\bar{K}=4$.
$t'_{1}=\{t_{1},t_{2},t_{3}\}$,$t'_{2}=\{t_{4},t_{5},t_{6},t_{7}\}$,
$t'_{3}=\{t_{8}\}$,$t'_{4}=\{t_{9},t_{10}\}$. To describe the
perfect Bayesian equilibrium of a signaling game with pattern
recognition, we give five basic requirements as follows:

R1: The receiver has a belief about which patterns can have sent
message $m$. These beliefs can be described as a probability
distribution $p(t'|m)$, the probability that the sender is of
pattern $t'$ if the sender chooses message $m$. The sum over all
patterns $t'\in T'$ of these probabilities has to be 1 conditional
on any message $m$.

R2: The receiver can observe the probability distribution of
messages, $p(m|t')$, when a pattern $t'$ is given.

R3: The action that the receiver chooses must maximize the expected
utility of the receiver given his beliefs about which patterns can
have sent message $m$. This means that the sum $\sum_{t'\in
T'}p(t'|m)\min\limits_{t\in t'}u_{2}(t,m,a)$ is maximized. The
action $a$ that maximizes this sum is denoted as $a^{*}(m,T')$. For
simplicity, we still use $a^{*}(m)$ to represent $a^{*}(m,T')$.

R4: For each type $t\in t'$ that the sender may have, the sender
chooses to send the message
\begin{equation*}
m^{*}(t')\in M^{*}(t')\equiv\{m|\max\limits_{m\in M, t\in
t'}u_{1}(t,m,a^{*}(m))\}
\end{equation*}
that maximizes the sender's
utility $u_{1}(t,m,a^{*}(m))$ given the strategy chosen by the
receiver $a^{*}(m)$.

R5: Let $|M^{*}(t')|$ be the cardinality of $M^{*}(t')$.
\begin{equation}
\widetilde{p}(m|t')=
\begin{cases}
   1/|M^{*}(t')|,& m\in M^{*}(t') \\
   0,& \mbox{otherwise}
\end{cases}
\end{equation}
Let $T'(m)$ denote the set of patterns that have incentives to send
$m$, i.e., $T'(m)=\{t'\in T'|m\in M^{*}(t')\}$. For each message $m$
that the sender can send, if there exists a type $t\in t'$ such that
$m^{*}(t')$ assigns strictly positive probability to $m$, the belief
$p(t'|m)$ that the receiver has about the pattern $t'$ of the sender
if he observes the message $m$ satisfies the Bayesian rule:
\begin{equation}
  \widetilde{p}(t'|m)=\frac{p(t')\widetilde{p}(m|t')}{\sum_{t'\in T'(m)}p(t')\widetilde{p}(m|t')}
\end{equation}

\textbf{Definition 2:} A perfect Bayesian equilibrium of a signaling
game with pattern recognition is composed of $m^{*}(t)$, $a^{*}(m)$,
$p(t'|m)$ and $p(m|t')$, which satisfies R1--R5.

\section{An example}
Consider a car trade game. There are two players in the market,
player 1 is the Seller, player 2 is the Buyer. The type space of
player 1 is $T=\{t_{1},t_{2},t_{3}\}$, where $t_{1},t_{2},t_{3}$
represent low, medium and high quality of his car respectively.
Player 1 knows his type but player 2 doesn't know it. The pattern
space hold by player 2 is $T'=\{t'_{1},t'_{2}\}$,
$t'_{1}=\{t_{1},t_{2}\}$, $t'_{2}=\{t_{3}\}$, i.e., in player 2's
mind, there are two patterns: ``Not high'' and ``High''. The belief
distribution hold by player 2 is $p(t'_{1})=\alpha$,
$p(t'_{2})=1-\alpha$.

The message space of player 1 is $M=\{m_{1},m_{2},m_{3}\}$, where
$m_{1}$, $m_{2}$ and $m_{3}$ represent three bids of the Seller,
$m_{3}>m_{2}>m_{1}>0$. The action space of player 2 is
$A=\{a_{1},a_{2}\}$, where $a_{1}$ stands for ``Not buy'', $a_{2}$
stands for ``Buy''. The Seller announces his bids to the Buyer, and
then the Buyer selects his action.

To describe the payoff functions of two players, $u_{1}(t,m,a)$ and
$u_{2}(t,m,a)$, we assume that the cost for the Seller of type
$t_{1}$ pretending to be of type $t_{2}$ is $c_{12}$. Similarly, we
define $c_{13}$ and $c_{23}$, and assume $c_{13}>c_{12}>0$,
$c_{13}>c_{23}>0$. The value for the Buyer when he buy a car of type
$t_{1},t_{2},t_{3}$ are $V_{1},V_{2},V_{3}$ respectively,
$V_{3}>V_{2}>V_{1}>0$.

1) When the Buyer selects $a_{1}$:
\begin{align*}
&u_{1}(t_{k},m_{j},a_{1})=0, k=1\sim3, j=1\sim3, j\leq k\\
&u_{1}(t_{1},m_{2},a_{1})=-c_{12}, u_{1}(t_{1},m_{3},a_{1})=-c_{13},
u_{1}(t_{2},m_{3},a_{1})=-c_{23}\\
&u_{2}(t_{k},m_{j},a_{1})=0, k=1\sim3, j=1\sim3
\end{align*}

2) When the Buyer selects $a_{2}$:
\begin{align*}
  &u_{1}(t_{1},m_{1},a_{2})=m_{1},
  &&u_{2}(t_{1},m_{1},a_{2})=V_{1}-m_{1}\\
  &u_{1}(t_{1},m_{2},a_{2})=m_{2}-c_{12},
  &&u_{2}(t_{1},m_{2},a_{2})=V_{1}-m_{2}\\
  &u_{1}(t_{1},m_{3},a_{2})=m_{3}-c_{13},
  &&u_{2}(t_{1},m_{3},a_{2})=V_{1}-m_{3}\\
  &u_{1}(t_{2},m_{1},a_{2})=m_{1},
  &&u_{2}(t_{2},m_{1},a_{2})=V_{2}-m_{1}\\
  &u_{1}(t_{2},m_{2},a_{2})=m_{2},
  &&u_{2}(t_{2},m_{2},a_{2})=V_{2}-m_{2}\\
  &u_{1}(t_{2},m_{3},a_{2})=m_{3}-c_{23},
  &&u_{2}(t_{2},m_{3},a_{2})=V_{2}-m_{3}\\
  &u_{1}(t_{3},m_{1},a_{2})=m_{1},
  &&u_{2}(t_{3},m_{1},a_{2})=V_{3}-m_{1}\\
  &u_{1}(t_{3},m_{2},a_{2})=m_{2},
  &&u_{2}(t_{3},m_{2},a_{2})=V_{3}-m_{2}\\
  &u_{1}(t_{3},m_{3},a_{2})=m_{3},
  &&u_{2}(t_{3},m_{3},a_{2})=V_{3}-m_{3}
\end{align*}

In order to obtain the perfect Bayesian equilibrium of the
second-hand car trade game, we show the requirements R1--R5 in
details:

\textbf{R1:} $\rho=p(t'_{1}|m_{1})$, $1-\rho=p(t'_{2}|m_{1})$.
$\sigma=p(t'_{1}|m_{2})$, $1-\sigma=p(t'_{2}|m_{2})$.
$\lambda=p(t'_{1}|m_{3})$, $1-\lambda=p(t'_{2}|m_{3})$.

\textbf{R2:} Player 2 can observe the probability $p(m_{j}|t'_{1})$
and $p(m_{j}|t'_{2})$, $j=1\sim3$.

\textbf{R3:} In order to computing the best strategy (action)
$a^{*}(m)$ of player 2 when he receives a message $m\in M$, we need
to solve the optimization problem:
\begin{equation}
  \max\limits_{a\in A}\sum\limits_{t'\in T'}p(t'|m)\min\limits_{t\in
  t'}u_{2}(t,m,a)
\end{equation}
For this example, the object function is represented as follows:
\begin{align*}
  &p(t'_{1}|m)\min\limits_{t\in t'_{1}}u_{2}(t,m,a)+p(t'_{2}|m)\min\limits_{t\in
  t'_{2}}u_{2}(t,m,a)\\
  =&p(t'_{1}|m)\min\{u_{2}(t_{1},m,a),u_{2}(t_{2},m,a)\}+p(t'_{2}|m)u_{2}(t_{3},m,a)
\end{align*}
If the Buyer selects $a_{1}$, the object function is 0. If the Buyer
selects $a_{2}$, the object function is:
\begin{equation}\label{eq2}  
  p(t'_{1}|m)u_{2}(t_{1},m,a)+p(t'_{2}|m)u_{2}(t_{3},m,a)
\end{equation}
1) If player 1 sends a message $m_{1}$, formula~\eqref{eq2} will be:
\begin{equation}\label{eq3}
\rho(V_{1}-m_{1})+(1-\rho)(V_{3}-m_{1})=(V_{3}-m_{1})-\rho(V_{3}-V_{1})
\end{equation}
Comparing formula~\eqref{eq3} with zero, there holds:
\begin{equation*}
  a^{*}(m_{1})=
\begin{cases}
   a_{1},& \rho_{0}\leq \rho \leq1 \\
   a_{2},& 0\leq\rho<\rho_{0}
\end{cases},\mbox{ where }\rho_{0}=\frac{V_{3}-m_{1}}{V_{3}-V_{1}}.
\end{equation*}
2) If player 1 sends a message $m_{2}$, formula~\eqref{eq2} will be:
\begin{equation}\label{eq4}
\sigma(V_{1}-m_{2})+(1-\sigma)(V_{3}-m_{2})=(V_{3}-m_{2})-\sigma(V_{3}-V_{1})
\end{equation}
Comparing formula~\eqref{eq4} with zero, there holds:
\begin{equation*}
  a^{*}(m_{2})=
\begin{cases}
   a_{1},& \sigma_{0}\leq \sigma \leq1 \\
   a_{2},& 0\leq\sigma<\sigma_{0}
\end{cases},\mbox{ where }\sigma_{0}=\frac{V_{3}-m_{2}}{V_{3}-V_{1}}.
\end{equation*}
3) If player 1 sends a message $m_{3}$, formula~\eqref{eq2} will be:
\begin{equation}\label{eq5}
\lambda(V_{1}-m_{3})+(1-\lambda)(V_{3}-m_{3})=(V_{3}-m_{3})-\lambda(V_{3}-V_{1})
\end{equation}
Comparing formula~\eqref{eq5} with zero, there holds:
\begin{equation*}
  a^{*}(m_{3})=
\begin{cases}
   a_{1},& \lambda_{0}\leq \lambda \leq1 \\
   a_{2},& 0\leq\lambda<\lambda_{0}
\end{cases},\mbox{ where }\lambda_{0}=\frac{V_{3}-m_{3}}{V_{3}-V_{1}}.
\end{equation*}
To sum the aforementioned optimal strategies of the Buyer, there
will be the following eight kinds of outcomes:

1) For $(\rho,\sigma,\lambda)\in D_{1}=\{\rho_{0}\leq \rho
\leq1,\sigma_{0}\leq \sigma \leq1,\lambda_{0}\leq \lambda
\leq1\}$,\\ $\quad a^{*}(m)=a_{1}$, $m\in M$.

2) For $(\rho,\sigma,\lambda)\in D_{2}=\{\rho_{0}\leq \rho
\leq1,\sigma_{0}\leq \sigma \leq1,0\leq \lambda <\lambda_{0}\}$,\\
$\quad a^{*}(m)=
\begin{cases}
   a_{1},& m=m_{1},m_{2} \\
   a_{2},& m=m_{3}
\end{cases}$.

3) For $(\rho,\sigma,\lambda)\in D_{3}=\{\rho_{0}\leq \rho
\leq1,0\leq\sigma<\sigma_{0},\lambda_{0}\leq \lambda\leq1\}$,\\
$\quad a^{*}(m)=
\begin{cases}
   a_{1},& m=m_{1},m_{3} \\
   a_{2},& m=m_{2}
\end{cases}$.

4) For $(\rho,\sigma,\lambda)\in D_{4}=\{\rho_{0}\leq \rho
\leq1,0\leq\sigma<\sigma_{0},0\leq \lambda <\lambda_{0}\}$,\\
$\quad a^{*}(m)=
\begin{cases}
   a_{1},& m=m_{1} \\
   a_{2},& m=m_{2},m_{3}
\end{cases}$.

5) For $(\rho,\sigma,\lambda)\in D_{5}=\{0\leq\rho<\rho_{0},
\sigma_{0}\leq \sigma \leq1,\lambda_{0}\leq \lambda \leq1\}$,\\
$\quad a^{*}(m)=
\begin{cases}
   a_{1},& m=m_{2},m_{3} \\
   a_{2},& m=m_{1}
\end{cases}$.

6) For $(\rho,\sigma,\lambda)\in D_{6}=\{0\leq\rho<\rho_{0},
\sigma_{0}\leq \sigma \leq1,0\leq \lambda <\lambda_{0}\}$,\\
$\quad a^{*}(m)=
\begin{cases}
   a_{1},& m=m_{2} \\
   a_{2},& m=m_{1},m_{3}
\end{cases}$.

7) For $(\rho,\sigma,\lambda)\in D_{7}=\{0\leq\rho<\rho_{0},
0\leq\sigma<\sigma_{0},\lambda_{0}\leq \lambda\leq1\}$,\\
$\quad a^{*}(m)=
\begin{cases}
   a_{1},& m=m_{3} \\
   a_{2},& m=m_{1},m_{2}
\end{cases}$.

8) For $(\rho,\sigma,\lambda)\in D_{8}=\{0\leq\rho<\rho_{0},
0\leq\sigma<\sigma_{0},0\leq \lambda <\lambda_{0}\}$,\\
$\quad a^{*}(m)=a_{2}$, $m\in M$.

\textbf{R4:} For each type $t$ ($t\in t'$) that the Seller may have,
we compute the optimal message $m^{*}(t')\in M^{*}(t')$ for the
Seller in these eight zones $D_{1},\cdots,D_{8}$, i.e., we need to
solve the maximization problem:
\begin{equation}\label{eq6}
\max\limits_{m\in M, t\in t'}u_{1}(t,m,a^{*}(m))
\end{equation}
1) Consider the zone $D_{1}$. There exists $a^{*}(m)=a_{1}$, $m\in
M$. \\For $t=t_{1}$,
\begin{align*}
&u_{1}(t_{1},m_{1},a^{*}(m_{1}))=u_{1}(t_{1},m_{1},a_{1})=0\\
&u_{1}(t_{1},m_{2},a^{*}(m_{2}))=u_{1}(t_{1},m_{2},a_{1})=-c_{12}\\
&u_{1}(t_{1},m_{3},a^{*}(m_{3}))=u_{1}(t_{1},m_{3},a_{1})=-c_{13}
\end{align*}
\begin{equation}\label{eq7}
m^{*}(t_{1})=m_{1}.
\end{equation}
For $t=t_{2}$,
\begin{align*}
&u_{1}(t_{2},m_{1},a^{*}(m_{1}))=u_{1}(t_{2},m_{1},a_{1})=0\\
&u_{1}(t_{2},m_{2},a^{*}(m_{2}))=u_{1}(t_{2},m_{2},a_{1})=0\\
&u_{1}(t_{2},m_{3},a^{*}(m_{3}))=u_{1}(t_{2},m_{3},a_{1})=-c_{23}
\end{align*}
\begin{equation}\label{eq8}
m^{*}(t_{2})=m_{1}\mbox{ or }m^{*}(t_{2})=m_{2}.
\end{equation}
For $t=t_{3}$,
\begin{align*}
&u_{1}(t_{3},m_{1},a^{*}(m_{1}))=u_{1}(t_{3},m_{1},a_{1})=0\\
&u_{1}(t_{3},m_{2},a^{*}(m_{2}))=u_{1}(t_{3},m_{2},a_{1})=0\\
&u_{1}(t_{3},m_{3},a^{*}(m_{3}))=u_{1}(t_{3},m_{3},a_{1})=0
\end{align*}
\begin{equation}\label{eq9}
m^{*}(t_{3})=m_{1}\mbox{ or }m^{*}(t_{3})=m_{2}\mbox{ or }
m^{*}(t_{3})=m_{3}.
\end{equation}
Comparing formula~\eqref{eq7},~\eqref{eq8} and~\eqref{eq9}, we
obtain the optimal message for the Seller: \\For
$(\rho,\sigma,\lambda)\in D_{1}$, $m^{*}(t)=m_{1}$, $t\in T$; or
$m^{*}(t)=\begin{cases}
   m_{1},& t=t_{1},t_{2} \\
   m_{2},& t=t_{3}
\end{cases}$;\\
or $m^{*}(t)=\begin{cases}
   m_{1},& t=t_{1},t_{3} \\
   m_{2},& t=t_{2}
\end{cases}$;
or $m^{*}(t)=\begin{cases}
   m_{1},& t=t_{1} \\
   m_{2},& t=t_{2},t_{3}
\end{cases}$;
or $m^{*}(t)=\begin{cases}
   m_{1},& t=t_{1} \\
   m_{2},& t=t_{2} \\
   m_{3},& t=t_{3}
\end{cases}$.
\\Therefore, for $(\rho,\sigma,\lambda)\in D_{1}$,
$M^{*}(t')=\{m_{1}\}$, $t'\in T'$; or $M^{*}(t')=\begin{cases}
   \{m_{1}\},& t'=t'_{1} \\
   \{m_{2}\},& t'=t'_{2}
\end{cases}$;\\
or $M^{*}(t')=\begin{cases}
   \{m_{1},m_{2}\},& t'=t'_{1} \\
   \{m_{1}\},& t'=t'_{2}
\end{cases}$;
or $M^{*}(t')=\begin{cases}
   \{m_{1},m_{2}\},& t'=t'_{1} \\
   \{m_{2}\},& t'=t'_{2}
\end{cases}$;\\
or $M^{*}(t')=\begin{cases}
   \{m_{1},m_{2}\},& t'=t'_{1} \\
   \{m_{3}\},& t'=t'_{2}
\end{cases}$.

2) Consider the zone $D_{2}$. There exists $a^{*}(m)=
\begin{cases}
   a_{1},& m=m_{1},m_{2} \\
   a_{2},& m=m_{3}
\end{cases}$.\\For $t=t_{1}$,
\begin{align*}
&u_{1}(t_{1},m_{1},a^{*}(m_{1}))=u_{1}(t_{1},m_{1},a_{1})=0\\
&u_{1}(t_{1},m_{2},a^{*}(m_{2}))=u_{1}(t_{1},m_{2},a_{1})=-c_{12}\\
&u_{1}(t_{1},m_{3},a^{*}(m_{3}))=u_{1}(t_{1},m_{3},a_{2})=m_{3}-c_{13}
\end{align*}
\begin{equation}\label{eq10}
m^{*}(t_{1})=\begin{cases}
   m_{1},& c_{13}\geq m_{3} \\
   m_{3},& 0<c_{13}<m_{3}
\end{cases}.
\end{equation}
For $t=t_{2}$,
\begin{align*}
&u_{1}(t_{2},m_{1},a^{*}(m_{1}))=u_{1}(t_{2},m_{1},a_{1})=0\\
&u_{1}(t_{2},m_{2},a^{*}(m_{2}))=u_{1}(t_{2},m_{2},a_{1})=0\\
&u_{1}(t_{2},m_{3},a^{*}(m_{3}))=u_{1}(t_{2},m_{3},a_{2})=m_{3}-c_{23}
\end{align*}
\begin{equation}\label{eq11}
m^{*}(t_{2})=\begin{cases}
   m_{1},& c_{23}\geq m_{3} \\
   m_{3},& 0<c_{23}<m_{3}
\end{cases},
\mbox{ or }m^{*}(t_{2})=\begin{cases}
   m_{2},& c_{23}\geq m_{3} \\
   m_{3},& 0<c_{23}<m_{3}
\end{cases}.
\end{equation}
For $t=t_{3}$,
\begin{align*}
&u_{1}(t_{3},m_{1},a^{*}(m_{1}))=u_{1}(t_{3},m_{1},a_{1})=0\\
&u_{1}(t_{3},m_{2},a^{*}(m_{2}))=u_{1}(t_{3},m_{2},a_{1})=0\\
&u_{1}(t_{3},m_{3},a^{*}(m_{3}))=u_{1}(t_{3},m_{3},a_{2})=m_{3}
\end{align*}
\begin{equation}\label{eq12}
m^{*}(t_{3})=m_{3}.
\end{equation}
Comparing formula~\eqref{eq10},~\eqref{eq11} and~\eqref{eq12}, we
obtain the optimal message  for the Seller: \\For
$(\rho,\sigma,\lambda)\in D_{2}$,

$\bullet$ If $0<c_{13}<m_{3}$ and $0<c_{23}<m_{3}$, then
$m^{*}(t)=m_{3}$, $t\in T$.

$\bullet$ If $0<c_{13}<m_{3}$ and $c_{23}\geq m_{3}$, then
$m^{*}(t)=\begin{cases}
   m_{1},& t=t_{2} \\
   m_{3},& t=t_{1},t_{3}
\end{cases}$,\\
or $m^{*}(t)=\begin{cases}
   m_{2},& t=t_{2} \\
   m_{3},& t=t_{1},t_{3}
\end{cases}$.

$\bullet$ If $c_{13}\geq m_{3}$ and $0<c_{23}<m_{3}$, then
$m^{*}(t)=\begin{cases}
   m_{1},& t=t_{1} \\
   m_{3},& t=t_{2},t_{3}
\end{cases}$.

$\bullet$ If $c_{13}\geq m_{3}$ and $c_{23}\geq m_{3}$, then
$m^{*}(t)=\begin{cases}
   m_{1},& t=t_{1},t_{2} \\
   m_{3},& t=t_{3}
\end{cases}$,\\
or $m^{*}(t)=\begin{cases}
   m_{1},& t=t_{1} \\
   m_{2},& t=t_{2} \\
   m_{3},& t=t_{3}
\end{cases}$.

Therefore, for $(\rho,\sigma,\lambda)\in D_{2}$,

$\bullet$ If $0<c_{13}<m_{3}$, then $M^{*}(t')=\{m_{3}\}$, $t'\in
T'$.

$\bullet$ If $c_{13}\geq m_{3}$ and $0<c_{23}<m_{3}$, then
$M^{*}(t')=\{m_{3}\}$, $t'\in T'$.

$\bullet$ If $c_{13}\geq m_{3}$ and $c_{23}\geq m_{3}$, then
$M^{*}(t')=\begin{cases}
   \{m_{1}\},& t'=t'_{1} \\
   \{m_{3}\},& t'=t'_{2}
\end{cases}$, or $M^{*}(t')=\begin{cases}
   \{m_{1}, m_{2}\},& t'=t'_{1} \\
   \{m_{3}\},& t'=t'_{2}
\end{cases}$.

3) Consider the zone $D_{3}$. There exists $a^{*}(m)=
\begin{cases}
   a_{1},& m=m_{1},m_{3} \\
   a_{2},& m=m_{2}
\end{cases}$.\\For $t=t_{1}$,
\begin{align*}
&u_{1}(t_{1},m_{1},a^{*}(m_{1}))=u_{1}(t_{1},m_{1},a_{1})=0\\
&u_{1}(t_{1},m_{2},a^{*}(m_{2}))=u_{1}(t_{1},m_{2},a_{2})=m_{2}-c_{12}\\
&u_{1}(t_{1},m_{3},a^{*}(m_{3}))=u_{1}(t_{1},m_{3},a_{1})=-c_{13}
\end{align*}
\begin{equation}\label{eq13}
m^{*}(t_{1})=\begin{cases}
   m_{1},& c_{12}\geq m_{2} \\
   m_{2},& 0<c_{12}<m_{2}
\end{cases}.
\end{equation}
For $t=t_{2}$,
\begin{align*}
&u_{1}(t_{2},m_{1},a^{*}(m_{1}))=u_{1}(t_{2},m_{1},a_{1})=0\\
&u_{1}(t_{2},m_{2},a^{*}(m_{2}))=u_{1}(t_{2},m_{2},a_{2})=m_{2}\\
&u_{1}(t_{2},m_{3},a^{*}(m_{3}))=u_{1}(t_{2},m_{3},a_{1})=-c_{23}
\end{align*}
\begin{equation}\label{eq14}
m^{*}(t_{2})=m_{2}.
\end{equation}
For $t=t_{3}$,
\begin{align*}
&u_{1}(t_{3},m_{1},a^{*}(m_{1}))=u_{1}(t_{3},m_{1},a_{1})=0\\
&u_{1}(t_{3},m_{2},a^{*}(m_{2}))=u_{1}(t_{3},m_{2},a_{2})=m_{2}\\
&u_{1}(t_{3},m_{3},a^{*}(m_{3}))=u_{1}(t_{3},m_{3},a_{1})=0
\end{align*}
\begin{equation}\label{eq15}
m^{*}(t_{3})=m_{2}.
\end{equation}
Comparing formula~\eqref{eq13},~\eqref{eq14} and~\eqref{eq15}, we
obtain the optimal message  for the Seller: \\For
$(\rho,\sigma,\lambda)\in D_{3}$,

$\bullet$ If $0<c_{12}<m_{2}$, then $m^{*}(t)=m_{2}$, $t\in T$.

$\bullet$ If $c_{12}\geq m_{2}$, then $m^{*}(t)=\begin{cases}
   m_{1},& t=t_{1} \\
   m_{2},& t=t_{2},t_{3}
\end{cases}$.

Therefore, for $(\rho,\sigma,\lambda)\in D_{3}$,
$M^{*}(t')=\{m_{2}\}$, $t'\in T'$.

4) Consider the zone $D_{4}$. There exists $a^{*}(m)=
\begin{cases}
   a_{1},& m=m_{1} \\
   a_{2},& m=m_{2},m_{3}
\end{cases}$.\\For $t=t_{1}$,
\begin{align*}
&u_{1}(t_{1},m_{1},a^{*}(m_{1}))=u_{1}(t_{1},m_{1},a_{1})=0\\
&u_{1}(t_{1},m_{2},a^{*}(m_{2}))=u_{1}(t_{1},m_{2},a_{2})=m_{2}-c_{12}\\
&u_{1}(t_{1},m_{3},a^{*}(m_{3}))=u_{1}(t_{1},m_{3},a_{2})=m_{3}-c_{13}
\end{align*}
\begin{equation}\label{eq16}
m^{*}(t_{1})=\begin{cases}
   m_{1},& c_{12}\geq m_{2}, c_{13}\geq m_{3} \\
   m_{2},& 0<c_{12}<m_{2}, c_{13}-c_{12}\geq m_{3}-m_{2} \\
   m_{3},& 0<c_{13}<m_{3}, c_{13}-c_{12}< m_{3}-m_{2}
\end{cases}.
\end{equation}
For $t=t_{2}$,
\begin{align*}
&u_{1}(t_{2},m_{1},a^{*}(m_{1}))=u_{1}(t_{2},m_{1},a_{1})=0\\
&u_{1}(t_{2},m_{2},a^{*}(m_{2}))=u_{1}(t_{2},m_{2},a_{2})=m_{2}\\
&u_{1}(t_{2},m_{3},a^{*}(m_{3}))=u_{1}(t_{2},m_{3},a_{2})=m_{3}-c_{23}
\end{align*}
\begin{equation}\label{eq17}
m^{*}(t_{2})=\begin{cases}
   m_{2},& c_{23}\geq m_{3}-m_{2} \\
   m_{3},& 0<c_{23}<m_{3}-m_{2}
\end{cases}.
\end{equation}
For $t=t_{3}$,
\begin{align*}
&u_{1}(t_{3},m_{1},a^{*}(m_{1}))=u_{1}(t_{3},m_{1},a_{1})=0\\
&u_{1}(t_{3},m_{2},a^{*}(m_{2}))=u_{1}(t_{3},m_{2},a_{2})=m_{2}\\
&u_{1}(t_{3},m_{3},a^{*}(m_{3}))=u_{1}(t_{3},m_{3},a_{2})=m_{3}
\end{align*}
\begin{equation}\label{eq18}
m^{*}(t_{3})=m_{3}.
\end{equation}
Comparing formula~\eqref{eq16},~\eqref{eq17} and~\eqref{eq18}, we
obtain the optimal message  for the Seller:\\For
$(\rho,\sigma,\lambda)\in D_{4}$,

$\bullet$ If $c_{12}\geq m_{2}$, $c_{13}\geq m_{3}$, $c_{23}\geq
m_{3}-m_{2}$, then \\$m^{*}(t)=\begin{cases}
   m_{1},& t=t_{1} \\
   m_{2},& t=t_{2} \\
   m_{3},& t=t_{3}
\end{cases}$. $M^{*}(t')=\begin{cases}
   \{m_{2}\},& t'=t'_{1} \\
   \{m_{3}\},& t'=t'_{2}
\end{cases}$.

$\bullet$ If $c_{12}\geq m_{2}$, $c_{13}\geq m_{3}$,
$0<c_{23}<m_{3}-m_{2}$, then \\$m^{*}(t)=\begin{cases}
   m_{1},& t=t_{1} \\
   m_{3},& t=t_{2}, t_{3}
\end{cases}$. $M^{*}(t')=\{m_{3}\}$, $t'\in T'$.

$\bullet$ If $0<c_{12}<m_{2}$, $c_{13}-c_{12}\geq m_{3}-m_{2}$,
$c_{23}\geq m_{3}-m_{2}$, then \\$m^{*}(t)=\begin{cases}
   m_{2},& t=t_{1}, t_{2} \\
   m_{3},& t=t_{3}
\end{cases}$. $M^{*}(t')=\begin{cases}
   \{m_{2}\},& t'=t'_{1} \\
   \{m_{3}\},& t'=t'_{2}
\end{cases}$.

$\bullet$ If $0<c_{12}<m_{2}$, $c_{13}-c_{12}\geq m_{3}-m_{2}$,
$0<c_{23}<m_{3}-m_{2}$, then \\$m^{*}(t)=\begin{cases}
   m_{2},& t=t_{1} \\
   m_{3},& t=t_{2}, t_{3}
\end{cases}$. $M^{*}(t')=\{m_{3}\}$, $t'\in T'$.

$\bullet$ If $0<c_{13}<m_{3}$, $c_{13}-c_{12}< m_{3}-m_{2}$,
$c_{23}\geq m_{3}-m_{2}$, then \\$m^{*}(t)=\begin{cases}
   m_{2},& t=t_{2} \\
   m_{3},& t=t_{1}, t_{3}
\end{cases}$. \\If $m_{3}-m_{2}\geq c_{13}$, then
$M^{*}(t')=\{m_{3}\}$, $t'\in T'$; else $M^{*}(t')=\begin{cases}
   \{m_{2}\},& t'=t'_{1} \\
   \{m_{3}\},& t'=t'_{2}
\end{cases}$.

$\bullet$ If $0<c_{13}<m_{3}$, $c_{13}-c_{12}< m_{3}-m_{2}$,
$0<c_{23}<m_{3}-m_{2}$, then \\$m^{*}(t)=m_{3}$, $t\in T$.
$M^{*}(t')=\{m_{3}\}$, $t'\in T'$.

5) Consider the zone $D_{5}$. There exists $a^{*}(m)=
\begin{cases}
   a_{1},& m=m_{2},m_{3} \\
   a_{2},& m=m_{1}
\end{cases}$.\\For $t=t_{1}$,
\begin{align*}
&u_{1}(t_{1},m_{1},a^{*}(m_{1}))=u_{1}(t_{1},m_{1},a_{2})=m_{1}\\
&u_{1}(t_{1},m_{2},a^{*}(m_{2}))=u_{1}(t_{1},m_{2},a_{1})=-c_{12}\\
&u_{1}(t_{1},m_{3},a^{*}(m_{3}))=u_{1}(t_{1},m_{3},a_{1})=-c_{13}
\end{align*}
\begin{equation}\label{eq19}
m^{*}(t_{1})=m_{1}.
\end{equation}
For $t=t_{2}$,
\begin{align*}
&u_{1}(t_{2},m_{1},a^{*}(m_{1}))=u_{1}(t_{2},m_{1},a_{2})=m_{1}\\
&u_{1}(t_{2},m_{2},a^{*}(m_{2}))=u_{1}(t_{2},m_{2},a_{1})=0\\
&u_{1}(t_{2},m_{3},a^{*}(m_{3}))=u_{1}(t_{2},m_{3},a_{1})=-c_{23}
\end{align*}
\begin{equation}\label{eq20}
m^{*}(t_{2})=m_{1}.
\end{equation}
For $t=t_{3}$,
\begin{align*}
&u_{1}(t_{3},m_{1},a^{*}(m_{1}))=u_{1}(t_{3},m_{1},a_{2})=m_{1}\\
&u_{1}(t_{3},m_{2},a^{*}(m_{2}))=u_{1}(t_{3},m_{2},a_{1})=0\\
&u_{1}(t_{3},m_{3},a^{*}(m_{3}))=u_{1}(t_{3},m_{3},a_{1})=0
\end{align*}
\begin{equation}\label{eq21}
m^{*}(t_{3})=m_{1}.
\end{equation}
Comparing formula~\eqref{eq19},~\eqref{eq20} and~\eqref{eq21}, we
obtain the optimal message  for the Seller: For
$(\rho,\sigma,\lambda)\in D_{5}$, $m^{*}(t)=m_{1}$, $t\in T$.
$M^{*}(t')=\{m_{1}\}$, $t'\in T'$.

6) Consider the zone $D_{6}$. There exists $a^{*}(m)=
\begin{cases}
   a_{1},& m=m_{2} \\
   a_{2},& m=m_{1},m_{3}
\end{cases}$.\\For $t=t_{1}$,
\begin{align*}
&u_{1}(t_{1},m_{1},a^{*}(m_{1}))=u_{1}(t_{1},m_{1},a_{2})=m_{1}\\
&u_{1}(t_{1},m_{2},a^{*}(m_{2}))=u_{1}(t_{1},m_{2},a_{1})=-c_{12}\\
&u_{1}(t_{1},m_{3},a^{*}(m_{3}))=u_{1}(t_{1},m_{3},a_{2})=m_{3}-c_{13}
\end{align*}
\begin{equation}\label{eq22}
m^{*}(t_{1})=\begin{cases}
   m_{1},& c_{13}\geq m_{3}-m_{1} \\
   m_{3},& 0<c_{13}<m_{3}-m_{1}
\end{cases}.
\end{equation}
For $t=t_{2}$,
\begin{align*}
&u_{1}(t_{2},m_{1},a^{*}(m_{1}))=u_{1}(t_{2},m_{1},a_{2})=m_{1}\\
&u_{1}(t_{2},m_{2},a^{*}(m_{2}))=u_{1}(t_{2},m_{2},a_{1})=0\\
&u_{1}(t_{2},m_{3},a^{*}(m_{3}))=u_{1}(t_{2},m_{3},a_{2})=m_{3}-c_{23}
\end{align*}
\begin{equation}\label{eq23}
m^{*}(t_{2})=\begin{cases}
   m_{1},& c_{23}\geq m_{3}-m_{1} \\
   m_{3},& 0<c_{23}<m_{3}-m_{1}
\end{cases}.
\end{equation}
For $t=t_{3}$,
\begin{align*}
&u_{1}(t_{3},m_{1},a^{*}(m_{1}))=u_{1}(t_{3},m_{1},a_{2})=m_{1}\\
&u_{1}(t_{3},m_{2},a^{*}(m_{2}))=u_{1}(t_{3},m_{2},a_{1})=0\\
&u_{1}(t_{3},m_{3},a^{*}(m_{3}))=u_{1}(t_{3},m_{3},a_{2})=m_{3}
\end{align*}
\begin{equation}\label{eq24}
m^{*}(t_{3})=m_{3}.
\end{equation}
Comparing formula~\eqref{eq22},~\eqref{eq23} and~\eqref{eq24}, we
obtain the optimal message for the Seller:\\For
$(\rho,\sigma,\lambda)\in D_{6}$,

$\bullet$ If $c_{23}\geq m_{3}-m_{1}$, then $m^{*}(t)=\begin{cases}
   m_{1},& t=t_{1},t_{2} \\
   m_{3},& t=t_{3}
\end{cases}$.

$\bullet$ If $0<c_{23}<m_{3}-m_{1}\leq c_{13}$, then
$m^{*}(t)=\begin{cases}
   m_{1},& t=t_{1} \\
   m_{3},& t=t_{2}, t_{3}
\end{cases}$.

$\bullet$ If $0<c_{13}<m_{3}-m_{1}$, then $m^{*}(t)=m_{3}$, $t\in
T$.

Therefore, for $(\rho,\sigma,\lambda)\in D_{6}$,

$\bullet$ If $c_{23}\geq m_{3}-m_{1}$, then $M^{*}(t')=\begin{cases}
   \{m_{1}\},& t'=t'_{1} \\
   \{m_{3}\},& t'=t'_{2}
\end{cases}$.

$\bullet$ If $0<c_{23}<m_{3}-m_{1}$, then $M^{*}(t')=\{m_{3}\}$,
$t'\in T'$.

7) Consider the zone $D_{7}$. There exists $a^{*}(m)=
\begin{cases}
   a_{1},& m=m_{3} \\
   a_{2},& m=m_{1},m_{2}
\end{cases}$.\\For $t=t_{1}$,
\begin{align*}
&u_{1}(t_{1},m_{1},a^{*}(m_{1}))=u_{1}(t_{1},m_{1},a_{2})=m_{1}\\
&u_{1}(t_{1},m_{2},a^{*}(m_{2}))=u_{1}(t_{1},m_{2},a_{2})=m_{2}-c_{12}\\
&u_{1}(t_{1},m_{3},a^{*}(m_{3}))=u_{1}(t_{1},m_{3},a_{1})=-c_{13}
\end{align*}
\begin{equation}\label{eq25}
m^{*}(t_{1})=\begin{cases}
   m_{1},& c_{12}\geq m_{2}-m_{1} \\
   m_{2},& 0<c_{12}<m_{2}-m_{1}
\end{cases}.
\end{equation}
For $t=t_{2}$,
\begin{align*}
&u_{1}(t_{2},m_{1},a^{*}(m_{1}))=u_{1}(t_{2},m_{1},a_{2})=m_{1}\\
&u_{1}(t_{2},m_{2},a^{*}(m_{2}))=u_{1}(t_{2},m_{2},a_{2})=m_{2}\\
&u_{1}(t_{2},m_{3},a^{*}(m_{3}))=u_{1}(t_{2},m_{3},a_{1})=-c_{23}
\end{align*}
\begin{equation}\label{eq26}
m^{*}(t_{2})=m_{2}.
\end{equation}
For $t=t_{3}$,
\begin{align*}
&u_{1}(t_{3},m_{1},a^{*}(m_{1}))=u_{1}(t_{3},m_{1},a_{2})=m_{1}\\
&u_{1}(t_{3},m_{2},a^{*}(m_{2}))=u_{1}(t_{3},m_{2},a_{2})=m_{2}\\
&u_{1}(t_{3},m_{3},a^{*}(m_{3}))=u_{1}(t_{3},m_{3},a_{1})=0
\end{align*}
\begin{equation}\label{eq27}
m^{*}(t_{3})=m_{2}.
\end{equation}
Comparing formula~\eqref{eq25},~\eqref{eq26} and~\eqref{eq27}, we
obtain the optimal message for the Seller:\\For
$(\rho,\sigma,\lambda)\in D_{7}$,

$\bullet$ If $c_{12}\geq m_{2}-m_{1}$, then $m^{*}(t)=\begin{cases}
   m_{1},& t=t_{1} \\
   m_{2},& t=t_{2}, t_{3}
\end{cases}$.\\
$\bullet$ If $0<c_{12}<m_{2}-m_{1}$, then $m^{*}(t)=m_{2}$, $t\in
T$.

Therefore, for $(\rho,\sigma,\lambda)\in D_{7}$,
$M^{*}(t')=\{m_{2}\}$, $t'\in T'$.

8) Consider the zone $D_{8}$. There exists $a^{*}(m)=a_{2}$, $m\in
M$.\\For $t=t_{1}$,
\begin{align*}
&u_{1}(t_{1},m_{1},a^{*}(m_{1}))=u_{1}(t_{1},m_{1},a_{2})=m_{1}\\
&u_{1}(t_{1},m_{2},a^{*}(m_{2}))=u_{1}(t_{1},m_{2},a_{2})=m_{2}-c_{12}\\
&u_{1}(t_{1},m_{3},a^{*}(m_{3}))=u_{1}(t_{1},m_{3},a_{2})=m_{3}-c_{13}
\end{align*}
\begin{equation}\label{eq28}
m^{*}(t_{1})=\begin{cases}
   m_{1},& c_{12}\geq m_{2}-m_{1}, c_{13}\geq m_{3}-m_{1} \\
   m_{2},& 0<c_{12}<m_{2}-m_{1}, c_{13}-c_{12}\geq m_{3}-m_{2} \\
   m_{3},& 0<c_{13}<m_{3}-m_{1}, c_{13}-c_{12}< m_{3}-m_{2}
\end{cases}.
\end{equation}
For $t=t_{2}$,
\begin{align*}
&u_{1}(t_{2},m_{1},a^{*}(m_{1}))=u_{1}(t_{2},m_{1},a_{2})=m_{1}\\
&u_{1}(t_{2},m_{2},a^{*}(m_{2}))=u_{1}(t_{2},m_{2},a_{2})=m_{2}\\
&u_{1}(t_{2},m_{3},a^{*}(m_{3}))=u_{1}(t_{2},m_{3},a_{2})=m_{3}-c_{23}
\end{align*}
\begin{equation}\label{eq29}
m^{*}(t_{2})=\begin{cases}
   m_{2},& c_{23}\geq m_{3}-m_{2} \\
   m_{3},& 0<c_{23}<m_{3}-m_{2}
\end{cases}.
\end{equation}
For $t=t_{3}$,
\begin{align*}
&u_{1}(t_{3},m_{1},a^{*}(m_{1}))=u_{1}(t_{3},m_{1},a_{2})=m_{1}\\
&u_{1}(t_{3},m_{2},a^{*}(m_{2}))=u_{1}(t_{3},m_{2},a_{2})=m_{2}\\
&u_{1}(t_{3},m_{3},a^{*}(m_{3}))=u_{1}(t_{3},m_{3},a_{2})=m_{3}
\end{align*}
\begin{equation}\label{eq30}
m^{*}(t_{3})=m_{3}.
\end{equation}
Comparing formula~\eqref{eq28},~\eqref{eq29} and~\eqref{eq30}, we
obtain the optimal message  for the Seller:\\For
$(\rho,\sigma,\lambda)\in D_{8}$,

$\bullet$ If $c_{12}\geq m_{2}-m_{1}$, $c_{13}\geq m_{3}-m_{1}$,
$c_{23}\geq m_{3}-m_{2}$, then \\$m^{*}(t)=\begin{cases}
   m_{1},& t=t_{1} \\
   m_{2},& t=t_{2} \\
   m_{3},& t=t_{3}
\end{cases}$. $M^{*}(t')=\begin{cases}
   \{m_{2}\},& t'=t'_{1} \\
   \{m_{3}\},& t'=t'_{2}
\end{cases}$.

$\bullet$ If $c_{12}\geq m_{2}-m_{1}$, $c_{13}\geq m_{3}-m_{1}$,
$0<c_{23}<m_{3}-m_{2}$, then \\$m^{*}(t)=\begin{cases}
   m_{1},& t=t_{1} \\
   m_{3},& t=t_{2}, t_{3}
\end{cases}$. $M^{*}(t')=\{m_{3}\}$, $t'\in T'$.

$\bullet$ If $0<c_{12}<m_{2}-m_{1}$, $c_{13}-c_{12}\geq
m_{3}-m_{2}$, $c_{23}\geq m_{3}-m_{2}$, then
\\$m^{*}(t)=\begin{cases}
   m_{2},& t=t_{1}, t_{2} \\
   m_{3},& t=t_{3}
\end{cases}$. $M^{*}(t')=\begin{cases}
   \{m_{2}\},& t'=t'_{1} \\
   \{m_{3}\},& t'=t'_{2}
\end{cases}$.

$\bullet$ If $0<c_{12}<m_{2}-m_{1}$, $c_{13}-c_{12}\geq
m_{3}-m_{2}$, $0<c_{23}<m_{3}-m_{2}$, then \\$m^{*}(t)=\begin{cases}
   m_{2},& t=t_{1} \\
   m_{3},& t=t_{2}, t_{3}
\end{cases}$. $M^{*}(t')=\{m_{3}\}$, $t'\in T'$.

$\bullet$ If $0<c_{13}<m_{3}-m_{1}$, $c_{13}-c_{12}< m_{3}-m_{2}$,
$c_{23}\geq m_{3}-m_{2}$, then \\$m^{*}(t)=\begin{cases}
   m_{2},& t=t_{2} \\
   m_{3},& t=t_{1}, t_{3}
\end{cases}$. $M^{*}(t')=\begin{cases}
   \{m_{2}\},& t'=t'_{1} \\
   \{m_{3}\},& t'=t'_{2}
\end{cases}$.

$\bullet$ If $0<c_{13}<m_{3}-m_{1}$, $c_{13}-c_{12}< m_{3}-m_{2}$,
$0<c_{23}<m_{3}-m_{2}$, then \\$m^{*}(t)=m_{3}$, $t\in T$.
$M^{*}(t')=\{m_{3}\}$, $t'\in T'$.

\textbf{R5:}\\For $M^{*}(t')=\{m_{1}\}$, $t'\in T'$:
\begin{align*}
  &T'(m_{1})=\{t'_{1},t'_{2}\}, T'(m_{2})=T'(m_{3})=\phi\\
  &\widetilde{p}(m_{1}|t')=1, \widetilde{p}(m_{2}|t')=\widetilde{p}(m_{3}|t')=0, t'\in
  T'\\
  &\widetilde{p}(t'_{1}|m_{1})=\frac{p(t'_{1})\widetilde{p}(m_{1}|t'_{1})}{p(t'_{1})\widetilde{p}(m_{1}|t'_{1})+p(t'_{2})\widetilde{p}(m_{1}|t'_{2})}=\alpha
\end{align*}
For $M^{*}(t')=\{m_{2}\}$, $t'\in T'$:
\begin{align*}
  &T'(m_{1})=T'(m_{3})=\phi, T'(m_{2})=\{t'_{1},t'_{2}\}\\
  &\widetilde{p}(m_{1}|t')=\widetilde{p}(m_{3}|t')=0, \widetilde{p}(m_{2}|t')=1, t'\in
  T'\\
  &\widetilde{p}(t'_{1}|m_{2})=\frac{p(t'_{1})\widetilde{p}(m_{2}|t'_{1})}{p(t'_{1})\widetilde{p}(m_{2}|t'_{1})+p(t'_{2})\widetilde{p}(m_{2}|t'_{2})}=\alpha
\end{align*}
For $M^{*}(t')=\{m_{3}\}$, $t'\in T'$:
\begin{align*}
  &T'(m_{1})=T'(m_{2})=\phi, T'(m_{3})=\{t'_{1},t'_{2}\}\\
  &\widetilde{p}(m_{1}|t')=\widetilde{p}(m_{2}|t')=0, \widetilde{p}(m_{3}|t')=1, t'\in
  T'\\
  &\widetilde{p}(t'_{1}|m_{3})=\frac{p(t'_{1})\widetilde{p}(m_{3}|t'_{1})}{p(t'_{1})\widetilde{p}(m_{3}|t'_{1})+p(t'_{2})\widetilde{p}(m_{3}|t'_{2})}=\alpha
\end{align*}
For $M^{*}(t')=\begin{cases}
   \{m_{1}\},& t'=t'_{1} \\
   \{m_{2}\},& t'=t'_{2}
\end{cases}$:
\begin{align*}
  &T'(m_{1})=\{t'_{1}\}, T'(m_{2})=\{t'_{2}\}, T'(m_{3})=\phi\\
  &\widetilde{p}(m_{1}|t'_{1})=\widetilde{p}(m_{2}|t'_{2})=1, \widetilde{p}(t'_{1}|m_{1})=\widetilde{p}(t'_{2}|m_{2})=1.
\end{align*}
For $M^{*}(t')=\begin{cases}
   \{m_{1}\},& t'=t'_{1} \\
   \{m_{3}\},& t'=t'_{2}
\end{cases}$:
\begin{align*}
  &T'(m_{1})=\{t'_{1}\}, T'(m_{2})=\phi, T'(m_{3})=\{t'_{2}\}\\
  &\widetilde{p}(m_{1}|t'_{1})=\widetilde{p}(m_{3}|t'_{2})=1, \widetilde{p}(t'_{1}|m_{1})=\widetilde{p}(t'_{2}|m_{3})=1.
\end{align*}
For $M^{*}(t')=\begin{cases}
   \{m_{2}\},& t'=t'_{1} \\
   \{m_{3}\},& t'=t'_{2}
\end{cases}$:
\begin{align*}
  &T'(m_{1})=\phi, T'(m_{2})=\{t'_{1}\}, T'(m_{3})=\{t'_{2}\}\\
  &\widetilde{p}(m_{2}|t'_{1})=\widetilde{p}(m_{3}|t'_{2})=1, \widetilde{p}(t'_{1}|m_{2})=\widetilde{p}(t'_{2}|m_{3})=1.
\end{align*}
For $M^{*}(t')=\begin{cases}
   \{m_{1},m_{2}\},& t'=t'_{1} \\
   \{m_{1}\},& t'=t'_{2}
\end{cases}$:
\begin{align*}
  &T'(m_{1})=\{t'_{1},t'_{2}\}, T'(m_{2})=\{t'_{1}\}, T'(m_{3})=\phi\\
  &\widetilde{p}(m_{1}|t'_{1})=\widetilde{p}(m_{2}|t'_{1})=0.5, \widetilde{p}(m_{1}|t'_{2})=1, \\
  &\widetilde{p}(t'_{1}|m_{1})=\frac{p(t'_{1})\widetilde{p}(m_{1}|t'_{1})}{p(t'_{1})\widetilde{p}(m_{1}|t'_{1})+p(t'_{2})\widetilde{p}(m_{1}|t'_{2})}=\frac{\alpha}{2-\alpha}\\
  &\widetilde{p}(t'_{2}|m_{1})=\frac{2-2\alpha}{2-\alpha}, \widetilde{p}(t'_{1}|m_{2})=1.
\end{align*}
For $M^{*}(t')=\begin{cases}
   \{m_{1},m_{2}\},& t'=t'_{1} \\
   \{m_{2}\},& t'=t'_{2}
\end{cases}$:
\begin{align*}
  &T'(m_{1})=\{t'_{1}\}, T'(m_{2})=\{t'_{1},t'_{2}\}, T'(m_{3})=\phi\\
  &\widetilde{p}(m_{1}|t'_{1})=\widetilde{p}(m_{2}|t'_{1})=0.5, \widetilde{p}(m_{2}|t'_{2})=1, \\
  &\widetilde{p}(t'_{1}|m_{2})=\frac{p(t'_{1})\widetilde{p}(m_{2}|t'_{1})}{p(t'_{1})\widetilde{p}(m_{2}|t'_{1})+p(t'_{2})\widetilde{p}(m_{2}|t'_{2})}=\frac{\alpha}{2-\alpha}.\\
  &\widetilde{p}(t'_{2}|m_{2})=\frac{2-2\alpha}{2-\alpha}, \widetilde{p}(t'_{1}|m_{1})=1.
\end{align*}
For $M^{*}(t')=\begin{cases}
   \{m_{1},m_{2}\},& t'=t'_{1} \\
   \{m_{3}\},& t'=t'_{2}
\end{cases}$:
\begin{align*}
  &T'(m_{1})=T'(m_{2})=\{t'_{1}\}, T'(m_{3})=\{t'_{2}\}\\
  &\widetilde{p}(m_{1}|t'_{1})=\widetilde{p}(m_{2}|t'_{1})=0.5, \widetilde{p}(m_{3}|t'_{2})=1, \\
  &\widetilde{p}(t'_{1}|m_{1})=\widetilde{p}(t'_{1}|m_{2})=\widetilde{p}(t'_{2}|m_{3})=1.
\end{align*}


\end{document}